\documentclass[prd,onecolumn,showpacs,floatfix,superscriptaddress,nofootinbib]{revtex4}
\usepackage{graphicx}
\usepackage{epsfig}
\usepackage{bm}
\usepackage{amssymb}
\usepackage{float}
\usepackage{amsmath}
\usepackage{dcolumn}
\usepackage{cancel}

\usepackage{mathrsfs}
\usepackage{dcolumn}
\usepackage{graphicx}
\usepackage{amsmath}
\usepackage{amsfonts}
\usepackage{amssymb}
\usepackage{microtype}
\usepackage{subfigure}
\usepackage{makeidx}
\usepackage{bm}
\usepackage{epsf}
\usepackage{color}
\usepackage{multirow,dcolumn}
\usepackage{graphicx}
\usepackage{mathrsfs}
\graphicspath{{Images/}}

\def\doi{http://doi.org}

\def\be{\begin{equation*}}
\def\ee{\end{equation*}}

\def\PR#1{{Phys.\ Rev.\ D \bf #1}}
\def\PRL#1{{Phys.\ Rev.\ Lett.\ \bf #1}}

\begin{document}

\title{Curvature Scalarization of Black Holes  in $f(R)$ Gravity }

\author{Zi-Yu Tang}
\email{tangziyu@sjtu.edu.cn}
\affiliation{Center for Astronomy and Astrophysics, School of Physics and Astronomy,
Shanghai Jiao Tong University, Shanghai 200240, China}
\affiliation{Collaborative Innovation Center of IFSA (CICIFSA),
Shanghai Jiao Tong University, Shanghai 200240, China}

\author{Bin Wang}
\email{wang\_b@sjtu.edu.cn}
\affiliation{School of Aeronautics and Astronautics, Shanghai Jiao Tong University, Shanghai 200240, China}
\affiliation{Collaborative Innovation Center of IFSA (CICIFSA),
Shanghai Jiao Tong University, Shanghai 200240, China}
\affiliation{Center for Gravitation and Cosmology, College of Physical Science
and Technology, Yangzhou University, Yangzhou 225009, China}

\author{Thanasis Karakasis}
\email{asis96kar@gmail.com} \affiliation{Physics Division,
National Technical University of Athens, 15780 Zografou Campus,
Athens, Greece.}

\author{Eleftherios Papantonopoulos}
\email{lpapa@central.ntua.gr} \affiliation{Physics Division,
National Technical University of Athens, 15780 Zografou Campus,
Athens, Greece.}

\vspace{3.5cm}

\begin{abstract}

We consider $f(R)$ gravity theories in the presence of a scalar field minimally coupled to gravity with a self-interacting potential. When the scalar field backreacts to the metric we find at large distances scalarized Schwarzschild-AdS and Schwarzschild-AdS-like black hole solutions. At small distances due to strong curvature effects and the scalar dynamis we find a rich structure of scalarized black hole solutions. When the scalar field is conformally coupled to gravity we also find  scalarized black hole solutions at small distances.
\end{abstract}

\maketitle

\flushbottom

\tableofcontents

\section{Introduction}

The study of black hole solutions with scalar hair is a very interesting aspect of General Relativity (GR) and had attracted a lot of interest.
 These hairy black holes are solutions generated from a modified Einstein-Hilbert action in which a scalar field  coupled to gravity is introduced.
However, these solutions in order the scalar field to be regular on the horizon and well behaved at large distances have to obey the powerful  no-hair theorems. The first hairy black hole solutions of GR were found in  asymptotically flat  spacetimes \cite{BBMB} but it was realized that the scalar field was divergent on the horizon and stability analysis shown that  these solutions were unstable \cite{bronnikov}. Therefore, a regularization procedure has to be found and make the scalar field regular on the horizon. The easiest way to comply with this requirement is to introduce a cosmological constant making  the scalar field regular on the horizon hiding all possible divergencies behind the horizon.

One of the first hairy black hole solution was discussed in \cite{Martinez:1996gn} which was a generalization
of the black hole solution in (2+1)-dimensions introduced in \cite{Banados:1992wn}. It is a static 3-dimensional black hole solution  circularly symmetric, asymptotically anti-de Sitter (AdS), with a scalar field conformably coupled to gravity with a scalar field  regular
everywhere. In 4-dimensions an exact black hole solution  with a negative cosmological constant and a minimally coupled
self-interacting scalar field was discussed in \cite{Martinez:2004nb}. The event horizon is a surface of negative
constant curvature enclosing the curvature singularity and the
spacetime is asymptotically locally AdS with a  scalar
field  regular everywhere.

Then various hairy black hole solutions
were found \cite{Zloshchastiev:2004ny}-\cite{Winstanley:2002jt}.  A
characteristic of these solutions is  that the parameters
connected with the scalar fields, the scalar charges, are connected in some way with
the  parameters of the hairy solution.  In \cite{Martinez:2005di} a topological black
hole dressed with a conformally coupled scalar field and
  electric charge was studied. Phase transitions of hairy topological black holes were
  studied in \cite{Koutsoumbas:2006xj,Martinez:2010ti}. An electrically charged black hole solution with a scalar
  field minimally coupled to gravity and electromagnetism
 was presented in \cite{Martinez:2006an}. It was found that regardless the value of the electric charge,
the black hole is massless and has a fixed temperature. The
thermodynamics of the solution was also studied. Further hairy
solutions  were reported in
\cite{Kolyvaris:2009pc}-\cite{Charmousis:2014zaa} with various
properties. More recently new hairy black hole solutions, boson
stars and numerical rotating hairy black hole solutions were
discussed
\cite{Dias:2011at,Stotyn:2011ns,Dias:2011tj,Kleihaus:2013tba}.
Also the thermodynamics of hairy black holes was studied in
\cite{Lu:2014maa}.

All these hairy black holes are solutions of an Einstein action of a GR theory in which the following terms are present. The Ricci scalar $R$, a length scale presented through the cosmological constant $\Lambda$, an electromagnetic field in the case of a charged hairy black holes and a scalar field which appears with its kinetic term minimally coupled to gravity and its potential. Then you solve the coupled Einstein-Maxwell-scalar field equations
assuming a spherically symmetric ansatz for the metric. The question is if you change or modify one of these basic ingredients of your action can you get regular exact hairy black hole solutions?

If the scalar field except its minimal coupling it is also coupled kinetically to Einstein tensor it was found in \cite{Koutsoumbas:2015ekk} that it takes more time for the black hole to be formed. If the  scalar field coupled to Einstein tensor backreacts to the metric new hairy black hole solutions can be generated. These gravity theories  belong to the general
scalar-tensor Horndeski theories. Then various hairy black hole solutions were found. In  \cite{Kolyvaris:2011fk,Kolyvaris:2013zfa} a gravity model
was considered consisting of an electromagnetic field and a scalar field coupled to the Einstein tensor
with vanishing cosmological constant. Then regular hairy black hole solutions were found evading the no-hair theorem thanks to the presence of the coupling of the scalar field to the Einstein tensor which plays the role of an effective cosmological constant.

In the context of the Hordenski theory there are also exact hairy black hole solutions. An exact Galileon black hole solution was analytically obtained in \cite{Rinaldi:2012vy} in a  static and spherically symmetric geometry. However, the scalar field which was coupled to Einstein tensor should be considered as a particle living outside the horizon of the black hole because it blows up on the horizon. Another exact Galileon black hole solution was discussed in \cite{Babichev:2013cya}.  It was found that for a static and spherically symmetric spacetime, the scalar field if it is time dependent, the
 no-hair arguments can be circumvented and a hairy black hole solution can be found. Recently there is a discussion of black holes with “soft hair” \cite{Hawking:2016msc}. These black holes do not carry scalar charge but soft gravitons or photons on their  horizons. Lifting the spherically symmetric requirement for the metric, it was found  that 3-dimensional and higher dimensional  Einstein gravity with negative cosmological constant admit stationary black holes with “soft hair” on the horizon \cite{Afshar:2016wfy, Grumiller:2019fmp}.

As we can see from the above discussion if we change the way the scalar field is coupled to gravity or the symmetric properties of the spacetime, we get hairy black hole solutions with quite different properties from the minimally coupled scalar field to gravity in spherically symmetric spacetimes. To the best of our knowledge, there is no any study of how the change of curvature effects the structure and properties of hairy black holes. Therefore, the aim of this work  is  to study if there are hairy black hole solutions and what are their properties in modified GR theories, in which except the linear Ricci term there are other
non-linear or even high order curvature terms. A particular class of models that includes higher order curvature invariants as functions of the Ricci scalar are the $f(R)$ gravity models. These theories were mainly introduced in an attempt to describe the early and late cosmological history of our Universe   \cite{DeFelice:2010aj}-\cite{Starobinsky:1980te}.  These theories  exclude contributions from any curvature invariants other than $R$ and they  avoid the Ostrogradski instability \cite{Ostrogradsky:1850fid} which usually is present in   higher derivative theories \cite{Woodard:2006nt}.

In $f(R)$ gravity theories there are black hole solutions which are simple deviations from the known black hole solutions of GR or they have completely new structures. Static spherically symmetric solutions in $f(R)$ gravity were studied in \cite{Multamaki:2006zb,Multamaki:2006ym}. Black hole solutions were investigated  with constant curvature,  with and without electric charge and cosmological constant in \cite{delaCruzDombriz:2009et,Hendi:2011eg,Sebastiani:2010kv}. Also various black hole solutions were found in $f(R)$ gravity thoeries having various properties \cite{Hollenstein:2008hp}-\cite{Nashed:2019tuk}. Recently in Maxwell-$f(R)$ gravity, a general exact charged black hole solution with dynamic curvature in  $D$-dimensions was found in \cite{Tang:2019qiy}. This general black hole solution can be reduced to the Reissner-Nordstr\"om (RN) black hole in $D$-dimensions in Einstein gravity and to the known charged black hole solutions with constant curvature in $f(R)$ gravity. Recently the stability of $f(R)$ black holes under scalar perturbations was studied in \cite{Aragon:2020xtm}.

In section \ref{scalarization} we present the scalarization procedure of black holes in $f(R)$ gravity.
In section \ref{conformalcoupling} we discuss the scalarization procedure in the case of scalar field conformally coupled to gravity. Finally in section \ref{conclusions} are a discussion and the conclusions.

\section{Scalarized Black Hole Solutions }
\label{scalarization}

It is known that the $f(R)$ gravity theories in the Einstein frame are equivalent to GR in the presence of a scalar field with a potential. Our approach in our study is to  consider the $f(R) $ gravity theory with a scalar field minimally coupled to gravity in the presence of a self-interacting potential. Varying this action
we will look for hairy black hole solutions. We will show that if this scalar field decouples, we recover $f(R)$ gravity.
 First we will consider the case without a self-interacting potential.

\subsubsection{Without self-interacting potential}

Consider the action
\begin{equation}
    S=\int d^4 x \sqrt{-g}\left\{\frac{1}{2\kappa}\left[f(R)-2\Lambda \right] -\frac{1}{2}g^{\mu\nu}\partial_\mu \phi \partial_\nu\phi  \right\}~,\label{zaction}
\end{equation}
where $\kappa$ is the Newton gravitational constant $\kappa=8\pi G$. The Einstein equations read
\begin{equation}
    f_R R_{\mu\nu}-\frac{1}{2}g_{\mu\nu}\left[f(R)-2\Lambda\right]+g_{\mu\nu}\square f_R-\nabla_\mu \nabla_\nu f_R=\kappa T_{\mu\nu}~,\label{EE}
\end{equation}
where $f_R \equiv f'(R)$ and the energy-momentum tensor $T_{\mu\nu}$ is given by
\begin{equation}
    T_{\mu\nu}=\partial_\mu \phi \partial_\nu\phi-\frac{1}{2}g_{\mu\nu}g^{\alpha\beta} \partial_\alpha \phi \partial_\beta\phi~. \label{T}
\end{equation}
The Klein-Gordon equation reads
\begin{equation}
    \square\phi=0~.
\end{equation}
We consider a spherically symmetric ansatz for the metric
\begin{equation}
    ds^2=-B(r)dt^2+\frac{1}{B(r)}dr^2+r^2d\theta^2+r^2\sin^2{\theta}d\varphi^2~. \label{metr}
\end{equation}
Then the Einstein equations become

\begin{itemize}

\item The $tt$ Einstein equation is

\begin{equation}
    \frac{B(r) \Big{(}r B'(r) f_R'(r)-f_R(r) \Big{(} r B''(r)+2 B'(r)\Big{)}+2 r B(r) f_R''(r)+4 B(r) f_R'(r)+\kappa  r B(r) \phi '(r)^2-r f(r)+2\Lambda  r\Big{)}}{r}=0~.
\end{equation}
\item The $rr$ Einstein equation is
\begin{equation}
   \frac{-r B'(r) f_R'(r)+f_R(r) \left(r B''(r)+2 B'(r)\right)-4 B(r) f_R'(r)+\kappa  r B(r) \phi '(r)^2+r f(r)-2\Lambda  r}{r B(r)}=0~.
\end{equation}
\item The $\theta\theta$  Einstein equation is
\begin{equation}
   2 f_R(r) \left(r B'(r)+B(r)-1\right)+r^2 f(r)=r \left(r \left(2 B'(r) f_R'(r)+2\Lambda \right)+B(r) \left(2 r f_R''(r)+2 f_R'(r)+\kappa  r \phi '(r)^2\right)\right)~,
\end{equation}
\item Finally the Klein-Gordon equation becomes

\begin{equation}
    B'(r) \phi '(r)+B(r) \left(\phi ''(r)+\frac{2 \phi '(r)}{r}\right)=0~.
\end{equation}

\end{itemize}

The Einstein equations $tt$ and $rr$ give a relation between $f_R(r)$ and $\phi(r)$
\begin{equation}
    \phi'(r)^2+f_R''(r)=0~, \label{eq01}
\end{equation}
while the Klein-Gordon equation gives a relation between $\phi(r)$ and $B(r)$ which it  can be written as
\begin{equation}
    \phi'(r)r^2 B(r)=C_1~,\label{nsol}
\end{equation}
where $C_1$ is an integration constant.

If we assume that there is a black hole solution with horizon $r_h$ then we must have $B(r_h)=0$. Then from relation (\ref{nsol}) we have $C_1=0$. However, this relation is valid for any $r$ which means that either $B(r)=0$ or $\phi'(r)=0$ should be zero. If $B(r)=0$ we do not have any geometry while if $\phi'(r)=0$  means that the scalar field is a constant everywhere. Therefore, we do not have any hairy black hole solution with a non-trivial scalar field.

\subsubsection{With self-interacting potential}

We have shown that if matter does not have self-interactions we can not have hairy black hole solutions. Therefore we further consider the $f(R) $ gravity theory with a scalar field minimally coupled in the presence of a self-interacting potential
\begin{equation}
    S=\int d^4 x \sqrt{-g}\left\{\frac{1}{2\kappa}\left[f(R) -2\Lambda \right] -\frac{1}{2}g^{\mu\nu}\partial_\mu \phi \partial_\nu\phi -V(\phi) \right\}~, \label{action1}
\end{equation}
where the scalar field and its self-interacting potential vanishes at space infinity
\begin{equation}
    \phi\left(r\to \infty\right)=0~,\quad V\left(r\to \infty\right)=0~,\quad V\big|_{\phi=0}=0~.
\end{equation}
Then the stress-energy tensor and the Klein-Gordon equation become
\begin{equation}
    T_{\mu\nu}=\partial_\mu \phi \partial_\nu\phi-g_{\mu\nu}\left[\frac{1}{2}g^{\alpha\beta} \partial_\alpha \phi \partial_\beta\phi+V(\phi)\right]~,
\end{equation}
\begin{equation}
    \square\phi=\frac{dV}{d\phi},\quad \frac{dV}{d\phi}\Big|_{\phi=0}=0~.
\end{equation}

Considering the metric ansatz (\ref{metr}) and setting $\kappa=1$ the field equations now become
\begin{equation}
    2 r B  f_R''-r f_R  B'' +r B'  f_R' -2 f_R  B'  +4 B  f_R' + r \left(B  \phi ' {}^2- f +2\Lambda  +2    V\right) =0~, \label{eq0}
\end{equation}
\begin{equation}
    r  f_R B''-r B'  f_R'+2  f_R B'-4 B  f_R'+r\left( B \phi '^2+f-2\Lambda  -2  V\right)=0~,\label{eq1}
\end{equation}
\begin{equation}
    2  f_R \left(r B'+B-1\right)=B r\left(2 r  f_R''+2  f_R'\right)+r^2 \left(2 B'  f_R'+B\phi '^2-f+2\Lambda+2V \right)~,\label{eq2}
\end{equation}
\begin{equation}
    B' \phi '+B \left(\phi ''+\frac{2 \phi '}{r}\right)=\frac{V'}{\phi '}~,\label{eqK}
\end{equation}
where the primes denote the derivatives with respect to $r$.

There are four equations (\ref{eq0}), (\ref{eq1}), (\ref{eq2}), and (\ref{eqK}), but only three of them are independent. We can use the first three of them to deduce the last one. In other words, we have four unknown quantities $B(r)$, $\phi(r)$, $f(R)$ and $V(\phi)$, while three independent equations. Therefore we need to choose one of these functions and then solve the others. Our initial motivation for this study was to see what is the effect of a matter distribution on a non-trivial curvature described by a $f(R)$ theory. Therefore we choose different distributions of matter $\phi(r)$ to see what kind of geometries, $f(R)$ theories and potentials can support such hairy structure.

Using the $t-t$ (\ref{eq0}) and $r-r$ (\ref{eq1}) components of the Einstein equations we recover the relation (\ref{eq01}) between $f_R$ and $\phi$ which is independent of the self-interaction of the scalar field, while the $t-t$ (\ref{eq0}) and $\theta-\theta$ (\ref{eq2}) components give the relation between $f_R$ and $B$
\begin{equation}
    f_R \left(r^2 B''-2 B+2\right)+r \left(r B'-2 B\right) f_R'=0 \label{eq02}~,
\end{equation}
from which we can solve $f_R(r)$ and $B(r)$
\begin{eqnarray}
 f_R(r)&=&c_1+c_2 r-\int\int \phi'{}^2drdr~,\label{eq999}\\
 B(r)&=&r^2 \left[\int \frac{-2\int{f_R}  dr +c_3}{ r^4 f_R} dr +c_4\right]~. \label{BBB}
\end{eqnarray}
We can see that if the scalar field $\phi(r)$ is known, then $f_R(r)$ can be obtained by integration and also  the metric function $B(r)$. The Klein-Gordon equation (\ref{eqK}) gives the expression of the potential,
\begin{equation}
    V(r)=\int \phi ' \left[B' \phi '+B \left(\phi ''+\frac{2 \phi '}{r}\right)\right] dr+V_0~, \label{V}
\end{equation}
which can also be obtained by integration.

Using (\ref{eq1}), we can obtain $f(r)$
\begin{equation}
    f(r)=B' f_R'-\frac{f_R \left(r B''+2 B'\right)}{r}+\frac{4 B f_R'}{r}-B \phi '^2+2 \Lambda +2 V~.
\end{equation}

Besides, the expression of curvature under our metric ansatz can be calculated through the metric function
\begin{equation}
    R(r)=-\frac{r^2 B''(r)+4 r B'(r)+2 B(r)-2}{r^2}~.
\end{equation}

From the expressions of $f(r)$, $R(r)$, $V(r)$ and $\phi(r)$ one can determine the $f(R)$ forms and the potentials $V(\phi)$.

In the action (\ref{action1}) we have introduced a cosmological constant $\Lambda$. However, in the expressions of the
functions $ f_R(r)$, $B(r)$, $V(r)$ and $ R(r)$ the cosmological constant does not appear. The reason is the presence of the
$f(R)$ function. However, as we will see in the next subsection an effective cosmological constant is generated.

We note here that the relation (\ref{eq01}) connects the $f(R)$ function with the scalar field $\phi$. This means that the $f(R)$ function has the information of the presence of the scalar charge. This equation leads to the equation (\ref{eq999}) in which if the scalar field $\phi$ decouples and $c_2=0$ we recover GR.

\subsection{Gaussian distribution}
and
We first consider the Gaussian distribution of the scalar field,
\begin{equation}
    \phi=A e^{-r^2/2}~, \label{gaus}
\end{equation}
where $A$ is the amplitude of the scalar field.

From (\ref{eq01}) we can solve the $f_R$ explicitly
\begin{equation}
    f_R(r)=-\frac{1}{4} A^2 \left(\sqrt{\pi } r E(r)+2 e^{-r^2}\right)+c_1+c_2 r~,
\end{equation}
where $c_1$, $c_2$ are integration constants and
\begin{equation}
    E(r)=\frac{2}{\sqrt{\pi}}\int_0^r e^{-r^2}dr
\end{equation}
is the Gauss error function.

In fact, we can use (\ref{BBB}) and (\ref{V}) to calculate the metric function
\begin{equation}
    B(r)=r^2 \left[\int \frac{2 r \left(A^2 e^{-r^2}-8 c_1-4 c_2 r\right)+ \sqrt{\pi } A^2 \left(2 r^2+3\right) E(r)+8c_3}{2r^4 \left(- A^2 \left(\sqrt{\pi } r E(r)+2 e^{-r^2}\right)+4c_1+4c_2 r\right)} dr+c_4\right]~,
\end{equation}
and the potential
\begin{equation}
    V(r)=V_0-A^2 \int e^{-r^2} r \left[\left(r^2-3\right) B-r B'\right] dr~.
\end{equation}

We rewrite the potential as a function of $\phi$,
\begin{equation}
    V(\phi)=V_0+\int \left[ B(\phi)\left(\ln{\frac{A^2}{\phi^2}}-3\right)+B'(\phi)\phi\ln{\frac{A^2}{\phi^2}}\right]\phi d\phi~,\label{eq11}
\end{equation}
where
\begin{eqnarray}
 B(\phi)&=&c_4\ln{\frac{A^2}{\phi^2}} -\ln{\frac{A^2}{\phi^2}} \int \frac{d\phi}{\phi f_R(\phi)} \left(\ln{\frac{A^2}{\phi^2}}\right)^{-5/2} \left[2\int \frac{f_R(\phi)d\phi}{\phi\sqrt{\ln{\frac{A^2}{\phi^2}}}} +c_3\right], \label{eq22)}\\
  f_R(\phi)&=&\frac{1}{2}\sqrt{\ln{\frac{A^2}{\phi^2}}}\int \frac{\phi d\phi}{\sqrt{\ln{\frac{A^2}{\phi^2}}}}-\frac{1}{2}\phi^2+c_1+c_2 \sqrt{\ln{\frac{A^2}{\phi^2}}}~.\label{eq33}
\end{eqnarray}

For the Gaussian distribution (\ref{gaus}) we have an exact solution for the metric function $B(\phi)$, the  $f(R)$ function and the potential $V(\phi)$ of the coupled field equations given by the equations (\ref{eq11})-(\ref{eq33}).

 If the scalar field is decoupled and $A=0$ then we have
\begin{eqnarray}
\phi(r)&=&0~,\\
f_R(r)&=&c_1+c_2 r~,\\
B(r)&=&1+\frac{c_2 c_3}{2 c_1^2}-\frac{c_3}{3 c_1 r}-\frac{c_2 r \left(c_1^2+c_2 c_3\right)}{c_1^3}+r^2 \left[c_4+\frac{c_2^2 \left(c_1^2+c_2 c_3\right) }{c_1^4}\ln \left(\frac{c_1}{r}+c_2\right)\right]~,\label{B}
\end{eqnarray}
and if we choose $c_2=0$ we get  the Schwarzschild-AdS black hole solution
\begin{equation}
    B(r)=1-\frac{c_3}{3 c_1 r}+c_4 r^2~.
\end{equation}
If $A \neq 0$ and the scalar field backreacts  with the metric we will study what kind of hairy black hole solutions we get and  what is their behaviour at large and small distances.

\subsubsection{Hairy black holes at large distances}

The asymptotic expressions at large $r$ distances are
\begin{eqnarray}
 B(r)&=&\frac{1}{2}-\frac{2M}{r}-\frac{r^2 \Lambda_{\text{eff} }}{3}+O\left(\frac{1}{r^2}\right)~, \label{eq44} \\
 V(r)&=&A^2 e^{-r^2} \left(\frac{1}{4} (2 \Lambda_{\text{eff} }+1) r^2-\frac{r^4\Lambda_{\text{eff} } }{6}-M r+O\left(r^0\right)\right)~,
\end{eqnarray}
where the parameter $M$ is related to the mass of black hole and $\Lambda_{\text{eff} }$ is an effective  cosmological constant
\begin{eqnarray}
M&=&\frac{4 c_1}{3 \sqrt{\pi } A^2-12 c_2}~,\label{M}\label{567}\\
\Lambda_{\text{eff} }&=&-3 c_4~,\label{568} \label{eq77}
\end{eqnarray}
where we had already adjusted the integration constant
\begin{equation}
    V_0=\frac{3}{2} \sqrt{\pi } A^2 M~,
\end{equation}
to make the potential vanish at $r$ infinity and it also satisfies
\begin{equation}
    \frac{dV(\phi)}{d\phi}\Big|_{\phi=0}=0~.
\end{equation}

The asymptotic expression of $f(r)$ at large $r$ distances is
\begin{equation}
    f(r)=2c_1 \left( \Lambda_{\text{eff} }+\frac{1}{r^2}-\frac{1}{3 M r}\right)+2 \Lambda+O\left(\frac{1}{r^3}\right)~.
\end{equation}

Note that at large $r$ distances  the curvature is
\begin{equation}
    R(r)=4 \Lambda_{\text{eff} }+\frac{1}{r^2}+O\left(\frac{1}{r^3}\right)~, \label{55}
\end{equation}
then we can obtain the form of the  $f(R)$ function
\begin{equation}
    f(R)\simeq c_1 \left(2R-6 \Lambda_{\text{eff} }-\frac{2}{3 M}\sqrt{R-4 \Lambda_{\text{eff} }}\right)+2 \Lambda~. \label{eq66}
\end{equation}

If we choose a specific value for the constant $c_2=\frac{\sqrt{\pi } A^2}{4}$ we get the approximate at large distances functions
\begin{eqnarray}
B(r)&=&1-\frac{2 M}{r}-\frac{\Lambda_{\text{eff}} }{3}r^2+O\left(\frac{1}{r^4}e^{-r^2}\right)~,\label{1234} \label{sca2}\\
R(r)&=&4\Lambda_{\text{eff}}+O\left(\frac{1}{r^2}e^{-r^2}\right)~,\label{sca4}\\
V(r)&=&A^2 e^{-r^2} \left(-\frac{\Lambda_{\text{eff}} }{6}r^4+\frac{\Lambda_{\text{eff}} }{2}r^2+\frac{r^2}{2}\right)+O\left(r e^{-r^2}\right)~,\\
f(r)&=& 2 (c_1\Lambda_{\text{eff}}+\Lambda )+O\left(r e^{-r^2}\right)~, \label{sca1}\\
V(\phi)&\simeq&-\phi^2\left[\frac{2\Lambda_{\text{eff}}}{3}\left(\ln{\frac{\phi}{A}}\right)^2+\left(\Lambda_{\text{eff}}+1\right)\ln{\frac{\phi}{A}}\right]~,\\
f(R)&\simeq & \frac{c_1}{2}R+2\Lambda~, \label{f(R)}
\end{eqnarray}
where
\begin{eqnarray}
M&=&\frac{\sqrt{\pi } A^2}{16 c_1}+\frac{c_3}{6 c_1}~,\label{sca3}\\
\Lambda_{\text{eff}}&=&-3c_4~. \label{eff1}
\end{eqnarray}

Let us summarize our results so far. In our explicit solutions of the field equations we have the four parameters $c_1, c_2, c_3, c_4$ and the scalar charge $A$. The parameter $c_4$ is related to the effective cosmological constant, the parameter $c_3$ is related to the mass M of the Schwarzschild-AdS
black hole while the parameters $c_1, c_2$ appear in the $f(R)$ function. We can see from (\ref{sca1}) that when we are at large distances the second term decouples and the $f(R)$ function goes to pure Ricci scalar term $R$. If we choose $c_2=\frac{\sqrt{\pi } A^2}{4}$ the scalar charge $A$ appears in the metric function (\ref{sca2}) though its mass (\ref{sca3}) scalarizing in this way the Schwarzschild-AdS black hole. Also from (\ref{sca4}) we can get  the usual relation $R(r)=4\Lambda_{\text{eff}}$. If we had chosen a different value of the constant $c_2$ we can see from relation (\ref{567}) and (\ref{568}) that we can generate other  Schwarzschild-AdS-like black hole solutions.
 Now the interesting question is if we go to small distances at which  the Ricci scalar is expected to get strong corrections and the scalar field to get stronger, what kind of scalarized black holes we can get?

\subsubsection{Hairy black holes at small distances}

The various  functions at origin $r\to 0$ can be expanded as
\begin{eqnarray}
B(r)&=&\frac{A^4-4 A^2 c_1+4 c_1^2+2 c_2 c_3}{\left(A^2-2 c_1\right)^2}+\frac{2 c_3}{\left(3 A^2 -6 c_1\right) r}+c_4 r^2+O\left(r^3\right)~,\\
R(r)&=&-\frac{4 c_2 c_3}{r^2 \left(A^2-2 c_1\right)^2}-12 c_4+O\left(r\right)~,\\
V(r)&=&V_1+\frac{4 A^2 c_3 r}{3 A^2-6 c_1}+\frac{3 A^2 r^2 \left(A^4-4 A^2 c_1+4 c_1^2+2 c_2 c_3\right)}{2 \left(A^2-2 c_1\right)^2}+O\left(r^3\right)~,\\
f(r)&=&\frac{2 c_2 c_3}{r^2 \left(A^2-2 c_1\right)}+\frac{4 c_2 \left(A^4-4 A^2c_1+4 c_1^2+2 c_2 c_3 \right)}{r \left(A^2-2 c_1\right)^2}+O\left(r^0\right)~,
\end{eqnarray}
where $V_1$ is an integration constant different from $V_0$. (New integration constant may comes out when we do the analysis, but that does not mean there are two free parameters. In fact, we can fix the numerical solutions by giving the boundary condition at one side $V(r\to \infty)=0$.)

When $A^2\neq 2c_1$ and $c_2\neq 0$, the curvature $R$ is divergent at origin $r\to 0$, indicating a singularity.

Note that the modified gravity $f(R)$ and its derivative $f'(R)$ need to satisfy the conditions
\begin{equation}
    \lim_{R\to \infty}\frac{f(R)-R}{R}=0~, \quad \lim_{R\to \infty}f'(R)-1=0~,
\end{equation}
which are necessary conditions to recover GR at early times to satisfy the restrictions from Big Bang nucleosynthesis and CMB, and at  high curvature regime for local system tests \cite{Pogosian:2007sw,Cembranos:2011sr}.

The two conditions give the same constraint
\begin{equation}
    A^2=2c_1-2~. \label{A}
\end{equation}
The functions $B(r), R(r), V(r), f(r)$ are all functions of the parameters $c_i$ and the scalar charge $A$. At large distances choosing various values of the $c_2$ parameter of the function $f(R)$ and with a non-zero scalar charge $A$ we get various scalarized black hole solutions. Choosing $c_2=\frac{\sqrt{\pi } A^2}{4}$ we saw that a scalarized Schwarzschild-AdS black hole is produced.

Using the  value of $c_2=\frac{\sqrt{\pi } A^2}{4}$, the effective cosmological constant (\ref{eff1}) and relation (\ref{A})
 we rewrite the asymptotic expressions at origin
\begin{eqnarray}
B(r)&=&1+\frac{1}{8} \sqrt{\pi } A^2 c_3-\frac{c_3}{3 r}-\frac{\Lambda_{\text{eff}} }{3}r^2+O\left(r^3\right)~,\\
R(r)&=&4 \Lambda_{\text{eff}}-\frac{\sqrt{\pi } A^2 c_3}{4 r^2}+O\left(r\right),\\
V(r)&=&\frac{3}{16} \sqrt{\pi } A^4 c_3 r^2-\frac{2}{3} A^2 c_3 r+\frac{3 A^2 r^2}{2}+V_1+O\left(r^3\right)~,\\
f(r)&=&\frac{\pi  A^4 c_3}{8 r}-\frac{\sqrt{\pi } A^2 c_3}{4 r^2}+\frac{\sqrt{\pi } A^2}{r}+2 \Lambda +2 \Lambda_{\text{eff}}+2V_1+O\left(r\right)~,
\end{eqnarray}
where
\begin{eqnarray}
V(\phi)&\simeq &V_1-\frac{1}{3} 2 \sqrt{2} A^{3/2} c_3 \sqrt{A-\phi }+\frac{3}{8} \sqrt{\pi } A^3 c_3 (A-\phi )+3 A (A-\phi )~,\\
f(R)&\simeq & R -2 \Lambda_{\text{eff}}+2 \Lambda+2V_1+\frac{\pi^{1/4} A \left(\sqrt{\pi } A^2 c_3+8\right) \sqrt{4 \Lambda_{\text{eff}}-R}}{4 \sqrt{c_3}}~.
\end{eqnarray}

If $c_3>0$ and $\Lambda_{\text{eff} }<0$, then
\begin{equation}
    B\left(r\to 0\right) \to -\frac{c_3}{3 r}\to -\infty, \quad  B\left(r\to \infty \right) \to -\frac{r^2 \Lambda_{\text{eff} }}{3} \to +\infty~.
\end{equation}

Using the constraint of $A$ (\ref{A}) we can show that  the metric function and also the  functions $V(r)$ and $f(r)$ are always continuous for any positive $r$. Therefore, there must exist a zero point, namely the event horizon of a black hole. In this case, the solution describes a scalarized black hole in AdS spacetimes as it can be seen in the following figures which are plotted  varying the mass M which is related to the $c_3$ parameter as follows, using (\ref{sca3}) and the relation  (\ref{A}) we have
\begin{equation}
    c_3=6 c_1 M-\frac{3 \sqrt{\pi } A^2}{8}=3 \left(A^2+2\right) M-\frac{3 \sqrt{\pi } A^2}{8}~.
\end{equation}

\begin{figure}[H]
    \centering
    \includegraphics[width=.40\textwidth]{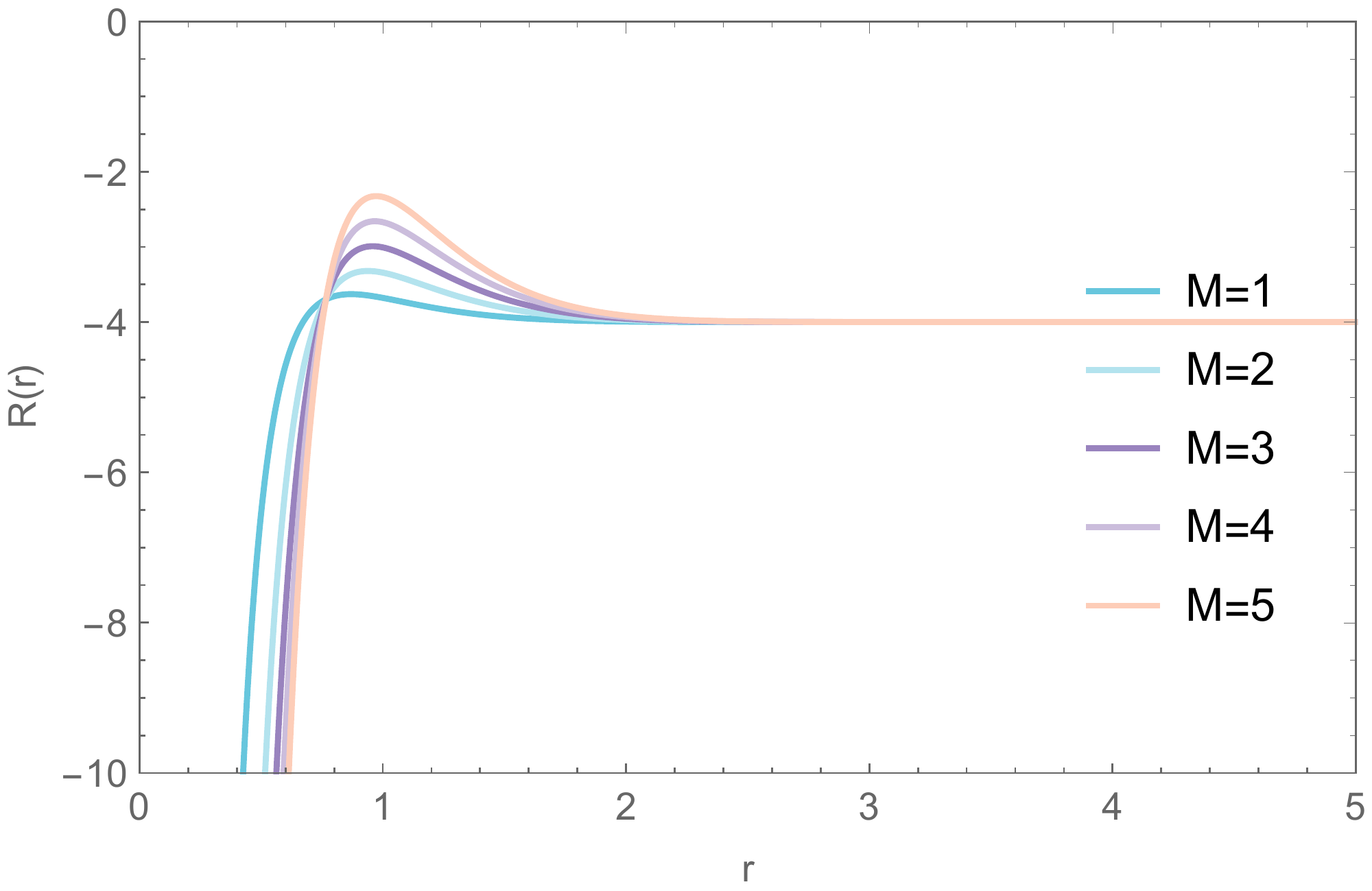}
     \includegraphics[width=.40\textwidth]{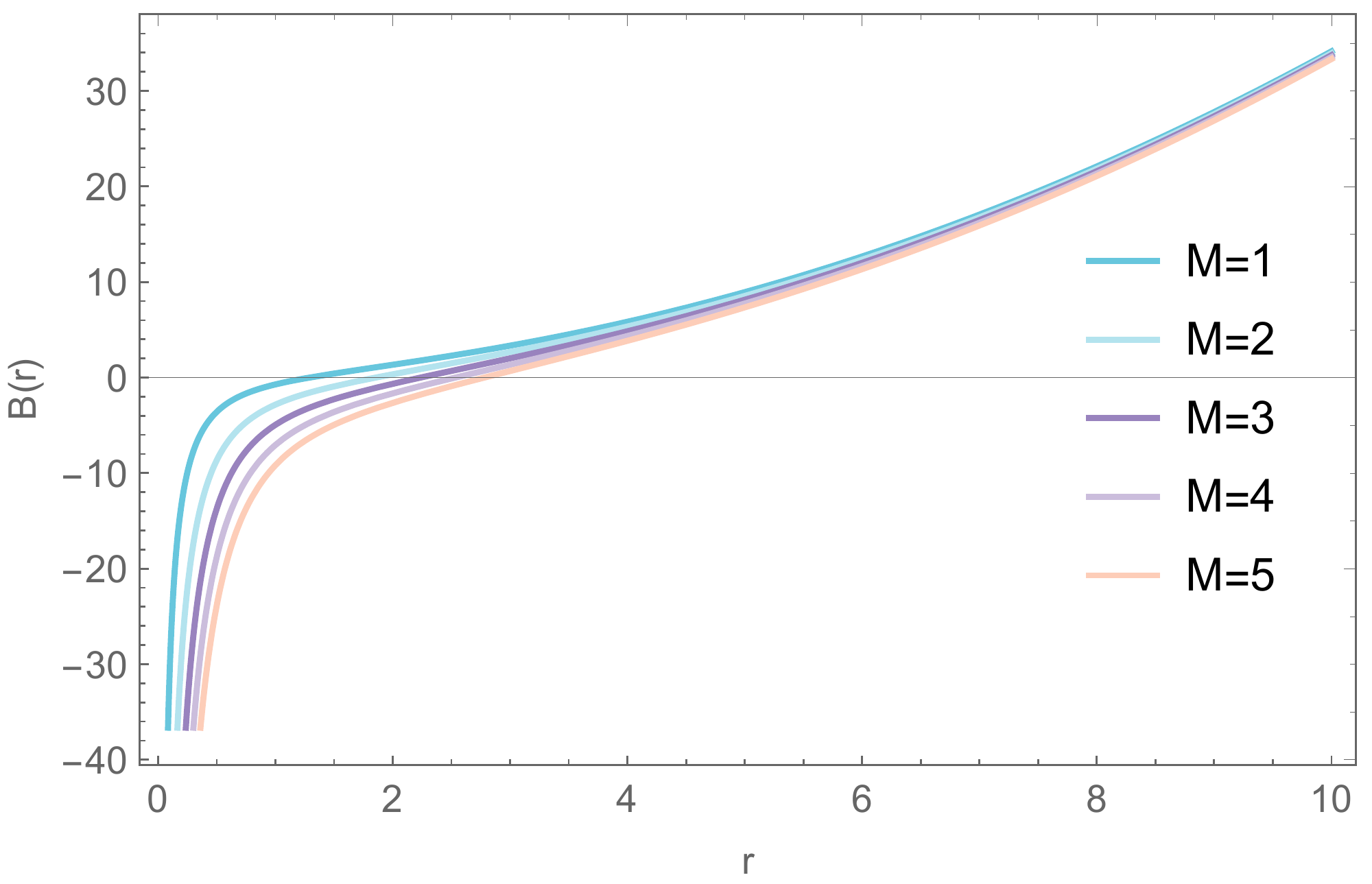}
    \caption{Plot of the curvature and the metric function for various values of $M$, while we have fixed $A=\sqrt{2}$ to have $c_1=2$ and $\Lambda_{\text{eff} }=-1$ .   } \label{fig11}
\end{figure}

\begin{figure}[H]
    \centering
    \includegraphics[width=.40\textwidth]{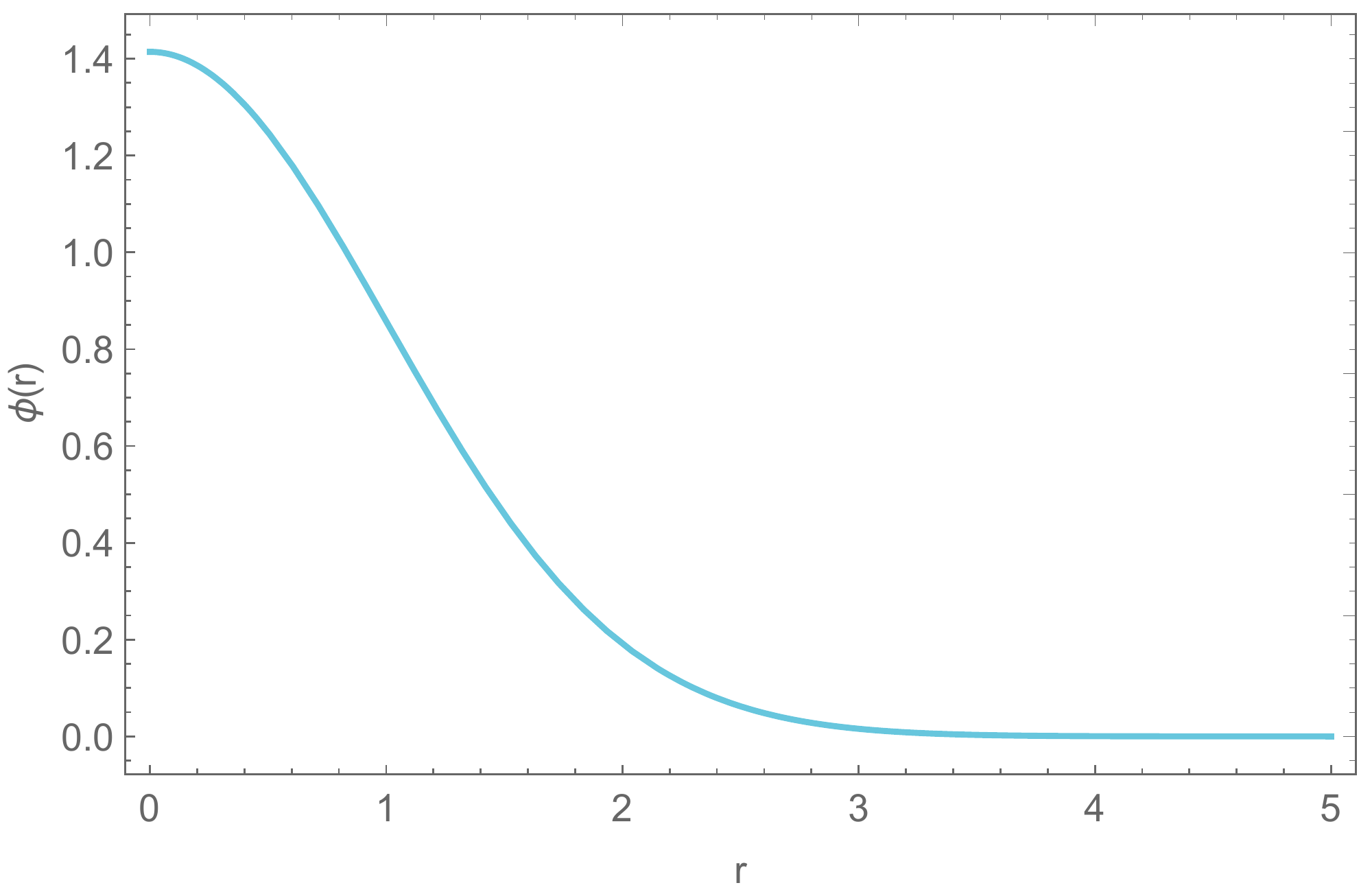}
     \includegraphics[width=.40\textwidth]{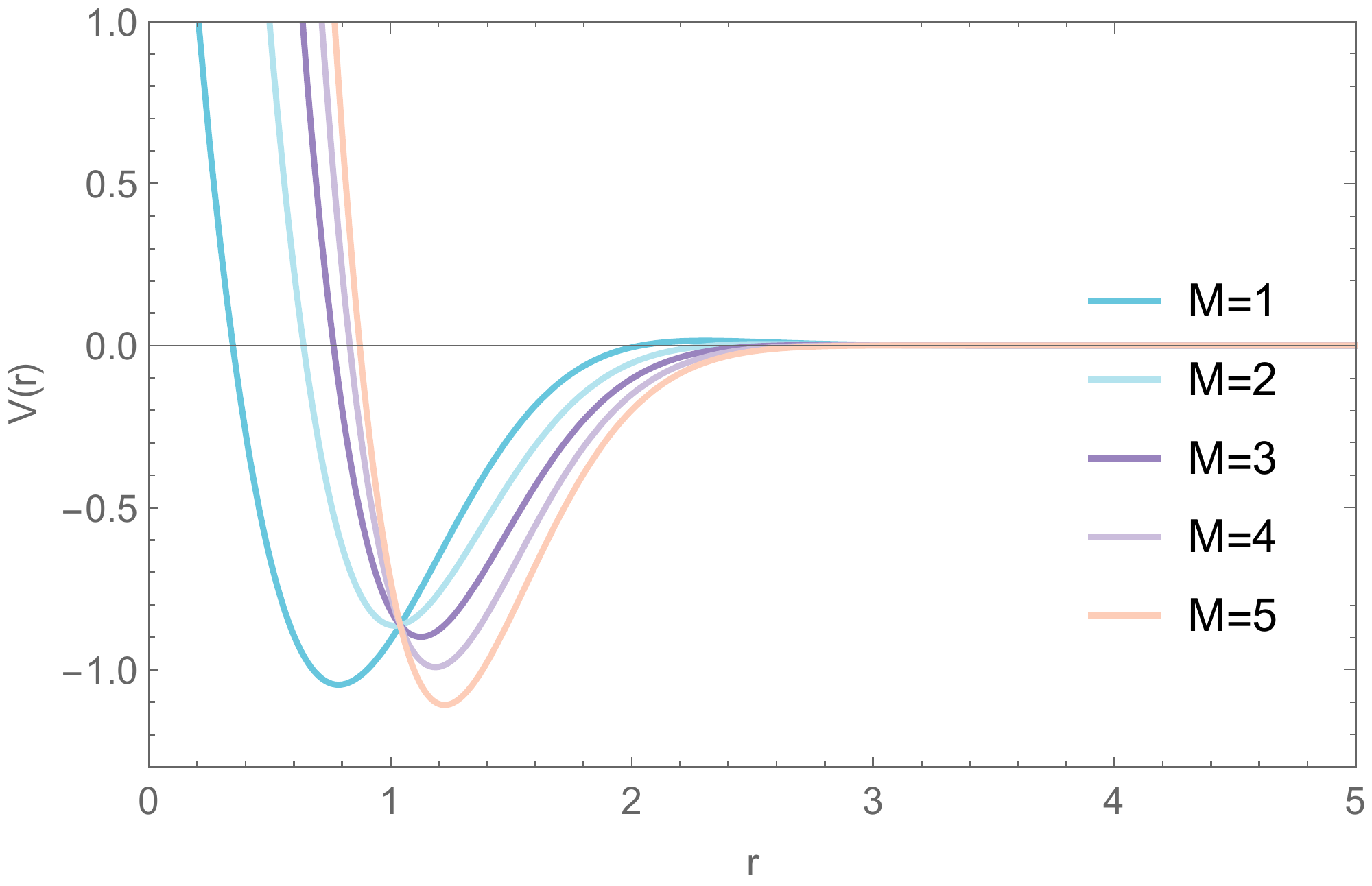}
    \caption{Plot of the scalar field and the potential  for various values  of $M$, while we have fixed $A=\sqrt{2}$ to have $c_1=2$ and $\Lambda_{\text{eff} }=-1$ .   } \label{fig22}
\end{figure}

\begin{figure}[H]
    \centering
    \includegraphics[width=.40\textwidth]{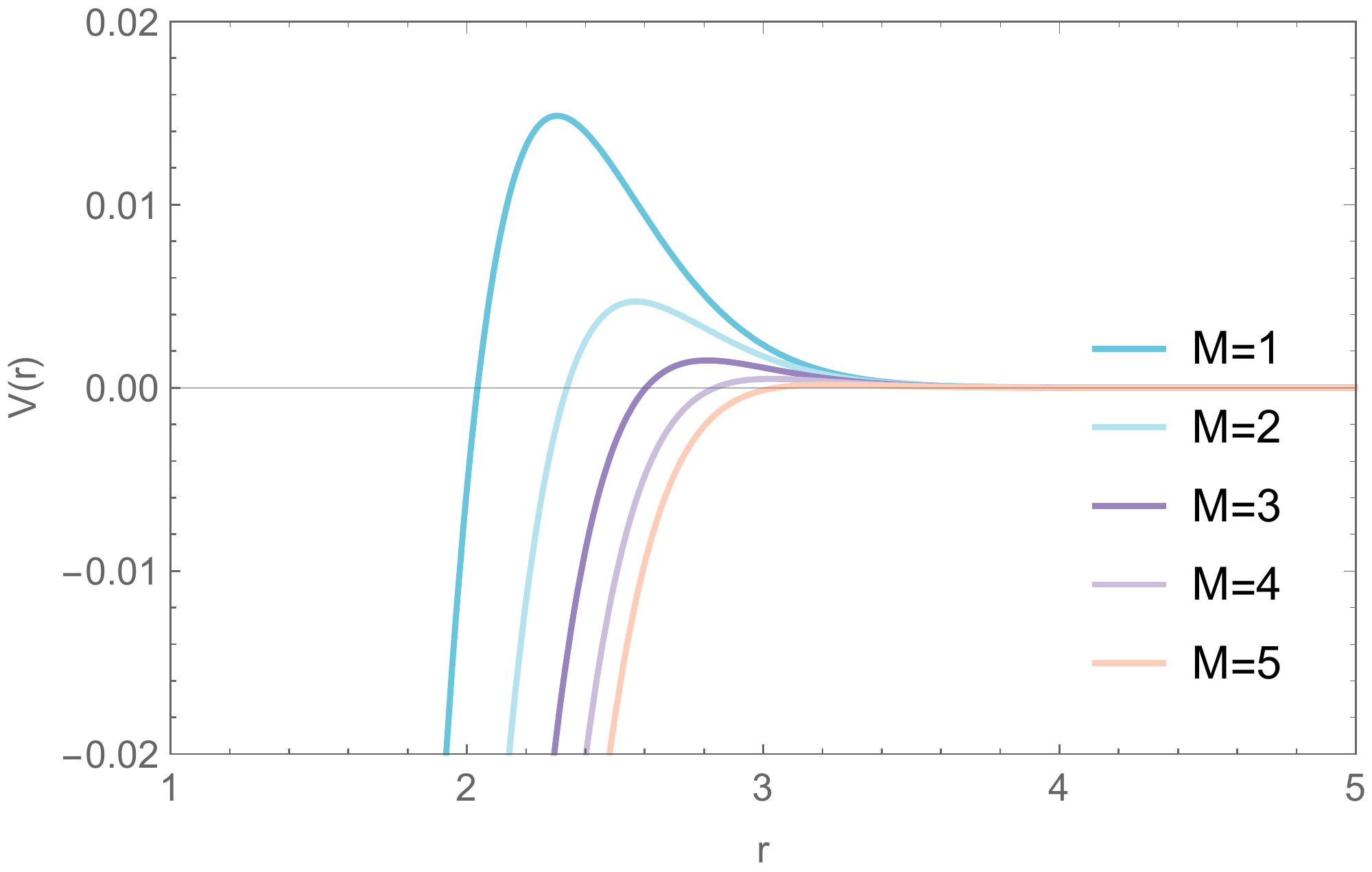}
    \caption{Plot of  the potential outside the horizon for various values  of $M$, while we have fixed $A=\sqrt{2}$ to have $c_1=2$ and $\Lambda_{\text{eff} }=-1$ .   } \label{fig222}
\end{figure}

The formation of a hairy black hole at small distances is very interesting. As can be seen in the left plot of Fig. \ref{fig11} the curvature blows up   at the origin while at the right plot the metric function develops  a horizon. In the left plot of Fig. \ref{fig22} the evolution of the scalar field is shown while the right plot  shows its potential. The potential develops a deep well. This well is formed before the appearance  of the horizon. This  indicates that the scalar field is trapped in this well providing the right matter concentration  for a hairy black hole to be formed. When the horizon is formed the potential of the scalar field develops a peak as it is shown in Fig. \ref{fig222}.

\begin{figure}[H]
    \centering
    \includegraphics[width=.40\textwidth]{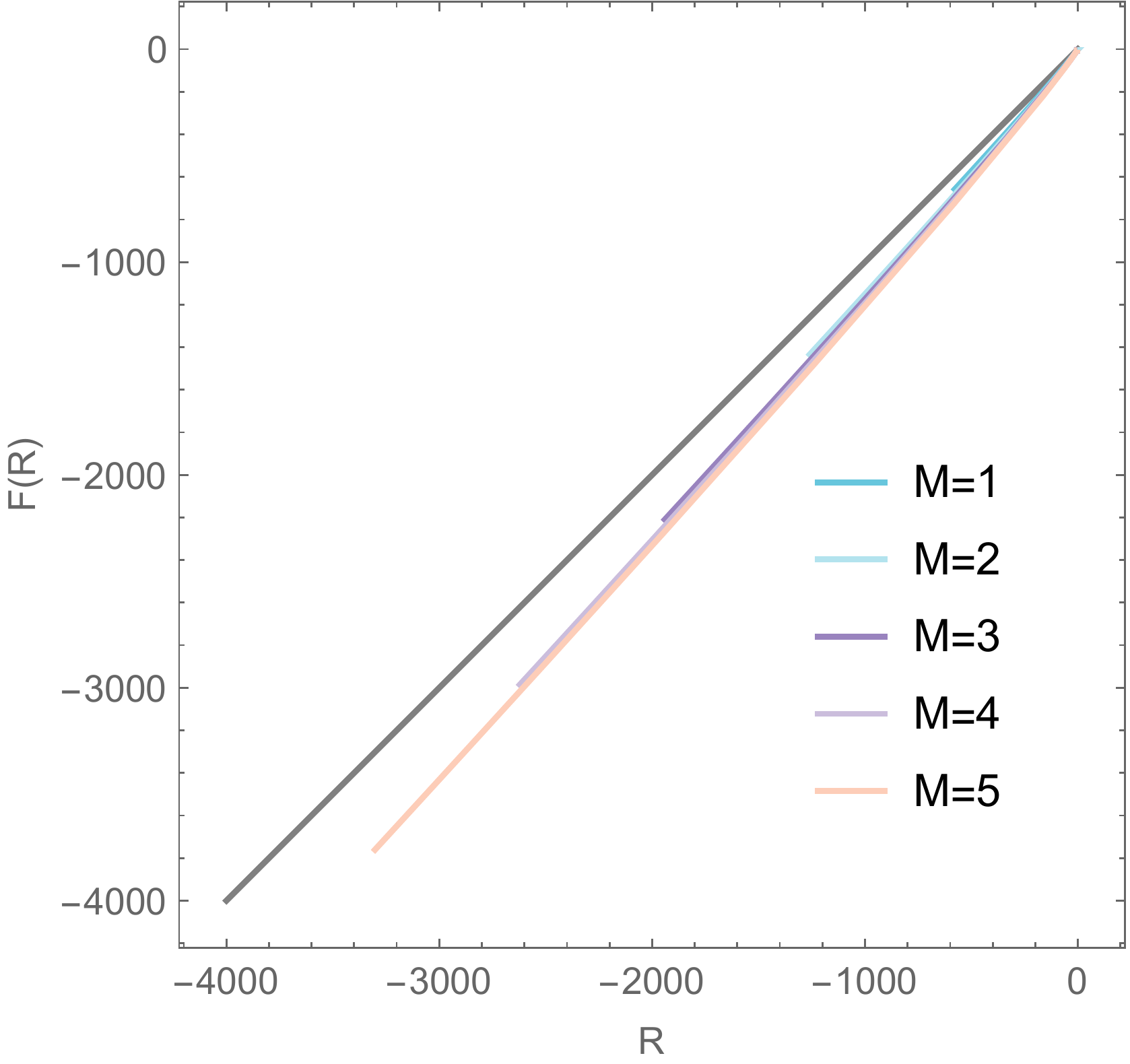}
    \caption{Compare the $F(R)$ functions with Einstein Gravity $F(R)=R$ (Gray line), while we have fixed $A=\sqrt{2}$ to have $c_1=2$ and $\Lambda_{\text{eff} }=-1$ .  } \label{fig33}
\end{figure}

Considering that we plot the figure with $\Lambda_{\text{eff}}$ instead of $\Lambda$, here we define a new function $F(R)$ which satisfies $F(R)-2\Lambda_{\text{eff}} \equiv f(R)-2\Lambda$. Finally in Fig. \ref{fig33} we compare this $F(R)$ function with Einstein Gravity $F(R)=R$. From the figure we can see the at small curvature it is very close to Einstein Gravity, while at large curvature it deviates from Einstein Gravity. In Einstein Gravity, such minimal coupling can not give hairy black holes due to no-hair theorems. While in our f(R) theory very close to Einstein Gravity as Fig. 4 shows, especially at small curvature, hairy black holes  can be obtained.

To have a better understating of  the  asymptotic structure near the event horizon $r\to r_h$, we make the following expansions of the metric, potential and modified gravity functions. These functions can be expanded as

\begin{eqnarray}
B(r)&=&B(r_h)+B'(r_h)(r-r_h)+\frac{1}{2}B''(r_h)(r-r_h)^2+...,\label{eq88}\\
V(r)&=&V(r_h)+V'(r_h)(r-r_h)+\frac{1}{2}V''(r_h)(r-r_h)^2+...,\label{eq99}\\
f(r)&=&f(r_h)+f'(r_h)(r-r_h)+\frac{1}{2}f''(r_h)(r-r_h)^2+....\label{eq10}
\end{eqnarray}


For simplicity we rewrite the metric function as
\begin{equation}
    B(r)=r^2 \left[\int b(r) dr+c_4\right]~, \label{eq101}
\end{equation}
where
\begin{equation}
    b(r)=\frac{2 r \left(A^2 e^{-r^2}-8 c_1-4 c_2 r\right)+ \sqrt{\pi } A^2 \left(2 r^2+3\right) E(r)+8c_3}{2r^4 \left(- A^2 \left(\sqrt{\pi } r E(r)+2 e^{-r^2}\right)+4c_1+4c_2 r\right)}~.
\end{equation}
Then
\begin{eqnarray}
B'(r)&=&\frac{B(r)}{r}+r^2 b(r)~,\\
B''(r)&=&\frac{2B(r)}{r^2}+4r b(r)+r^2 b'(r)~,\\
B'''(r)&=& r^2 b''(r)+6r b'(r)+6 b(r)~,\\
&...&
\end{eqnarray}
and at the event horizon $r_h$ we have
\begin{eqnarray}
B(r_h)&=&0~,\\
B'(r_h)&=&r_h^2 b(r_h)~,\\
B''(r_h)&=&4r_h b(r_h)+r_h^2 b'(r_h)~,\\
B'''(r_h)&=&r_h^2 b''(r_h)+6r_h b'(r_h)+6 b(r_h)~,\\
&...&
\end{eqnarray}

For the potential $V(r)$ we have
\begin{eqnarray}
V'(r_h)&=&B'(r_h) \phi '(r_h)^2=r_h^2 b(r_h) \phi '(r_h)^2~,
\end{eqnarray}
and also
\begin{equation}
    f'(r_h)=b(r_h) \left(r_h \left(r_h \left(f_R''(r_h)+\phi '(r_h)^2\right)+2 f_R'(r_h)\right)-12 f_R(r_h)\right)-r_h f_R(r_h) \left(r_h b''(r_h)+8 b'(r_h)\right)~.
\end{equation}

As can be seen from the expressions  (\ref{eq88})-(\ref{eq10}) around the horizon of the metric function $B(r)$, the potential $V(r)$ and the function $f(r)$, if the scalar field backreacts to the metric at small distances there is no any exact hairy black hole solution and a numerically hairy black hole solution is resulted from a metric function (\ref{eq101}) which is not simple. This can be understood from the fact that at small distances the modified curvature of the $f(R)$ theory is so strong that it gives strongly coupled hairy black holes.

\subsection{Other matter distributions}

If we consider another matter distribution we expect to get a similar structure of the hairy black holes at large and small distances. This will depend on the behaviour of the scalar field at large and small distances. Choosing different matter distributions will affect the form of the $f(R)$ function which  nevertheless it will always  have extra curvature terms other than the Ricci scalar $R$. If for example we choose a a polynomial distribution like
\begin{equation}
    \phi(r)=\frac{A}{(r+s)^p}~,
\end{equation}
we will get
\begin{equation}
    f_R(r)=-\frac{A^2 p (r+s)^{-2 p}}{4 p+2}+c_1+c_2 r~,
\end{equation}
or an inverse trigonometric function distribution of the scalar field
\begin{equation}
    \phi(r)=\frac{\pi }{2}-\arctan{r}~,
\end{equation}
we will get
\begin{equation}
    f_R(r)=-\frac{1}{2} r \arctan{r}+c_1+c_2 r~.
\end{equation}

Therefore we expect that strong and weak curvature effects of the $f(R)$ function at small and large distances will give hairy black hole solutions.

\section{Non Minimal Coupling}
\label{conformalcoupling}

In this section we will consider a scalar field non minimally coupled to gravity and we will look for hairy black holes.
The strength of the coupling between the scalar field and gravity is denoted by the factor $1/12$ (the conformal coupling factor) and in the action we also consider a self interacting potential. The motivation for this study is to show that   choosing  various matter distributions  we can have the formation of hairy black holes at small distances if the scalar field is not minimally coupled. We will see that the scalarization mechanism depends internally on the dynamics of the scalar field before and after the formation of the horizon of the black hole.

Consider the action
\begin{equation}
S= \int d^{4}x \sqrt{-g}\big[\cfrac{1}{2 \kappa }\big(f(R)-2\Lambda \big) - \cfrac{1}{2}g^{\mu\nu}\partial_{\mu}\phi\partial_{\nu}\phi - \cfrac{1}{12}R\phi^{2} -V(\phi)\big]~.
\end{equation}
Varying this action we get the same field equation (\ref{EE}) but with a different energy-momentum tensor
\begin{equation} T_{\mu \nu} = \partial_{\mu}\phi\partial_{\nu}\phi - \cfrac{1}{2}g_{\mu \nu}g^{\alpha \beta}\partial_{\alpha}\phi\partial_{\beta}\phi + \cfrac{1}{6} \left(g_{\mu \nu}\Box - \nabla_{\mu}\nabla_{\nu} + G_{\mu \nu}\right)\phi^{2} -g_{\mu \nu}V(\phi)~, \end{equation}
and a different Klein-Gordon equation
\begin{equation} \Box \phi -\cfrac{1}{6}R\phi - V'(\phi)=0~.
\end{equation}

Using same metric ansatz (\ref{metr}) and setting $\kappa=1$, from the $\theta \theta $ and $ tt$  components of the Einstein equations we get
\begin{equation}\cfrac{f_R (r) \left(-3 r^2 B''(r)+6 B(r)-6\right)+r \left(r B'(r)-2 B(r)\right) \left(\phi (r) \phi '(r)-3 f_R '(r)\right)}{12 r^2 B(r)} =0~,\label{ththtt} \end{equation}
while from the $tt$ and $rr$  components of the Einstein equations we get
\begin{equation}\frac{1}{6} B(r)^2 \left(3 f_R''(r)-\phi (r) \phi ''(r)+2 \phi '(r)^2\right)=0~.\label{ttrr}\end{equation}
The Klein-Gordon equation reads
\begin{equation}\frac{\left(r B'(r)+2 B(r)\right) \phi '(r)}{r}+\frac{\phi (r) \left(r^2 B''(r)+4 r B'(r)+2 B(r)-2\right)}{6 r^2}+B(r) \phi ''(r)-\frac{V'(r)}{\phi '(r)}=0~.\label{klein1} \end{equation}
As in the previous section with a scalar field minimally coupled  to gravity, we will consider various matter distributions and we will study their effect on a spherically symmetric metric. Then having the  forms of the scalar field $\phi(r)$  we will solve for $f_{R}(r)$ from equation (\ref{ttrr}), then we will get  $B(r)$ and $V(r)$ numerically from equations (\ref{ththtt}) and (\ref{klein1}) respectively. 

In this case, the equations are hard to be even asymptotically integrated in full generality. Thus we can not give a proof of continuity like we did in the previous section. But for the boundary conditions we give, the numerical results show that they are continuous, then an horizon is formed, and we indeed observe similar behaviors with the minimally coupled case.

\begin{subsection}{ $\phi(r) = \cfrac{m}{r+n}$}

We first consider the form of the scalar field $\phi(r) = \cfrac{m}{r+n}$ where $m$ and $n$ are constants. Then we get
\begin{equation} f_{R} = C_1 + C_2r~,\label{R1} \end{equation}
and the metric function $B(r)$, $V(r)$ can be obtained numerically as shown in Fig. \ref{fig1} and Fig. \ref{fig2}.
\begin{figure}[h]
\includegraphics[width = 8cm, height = 4cm]{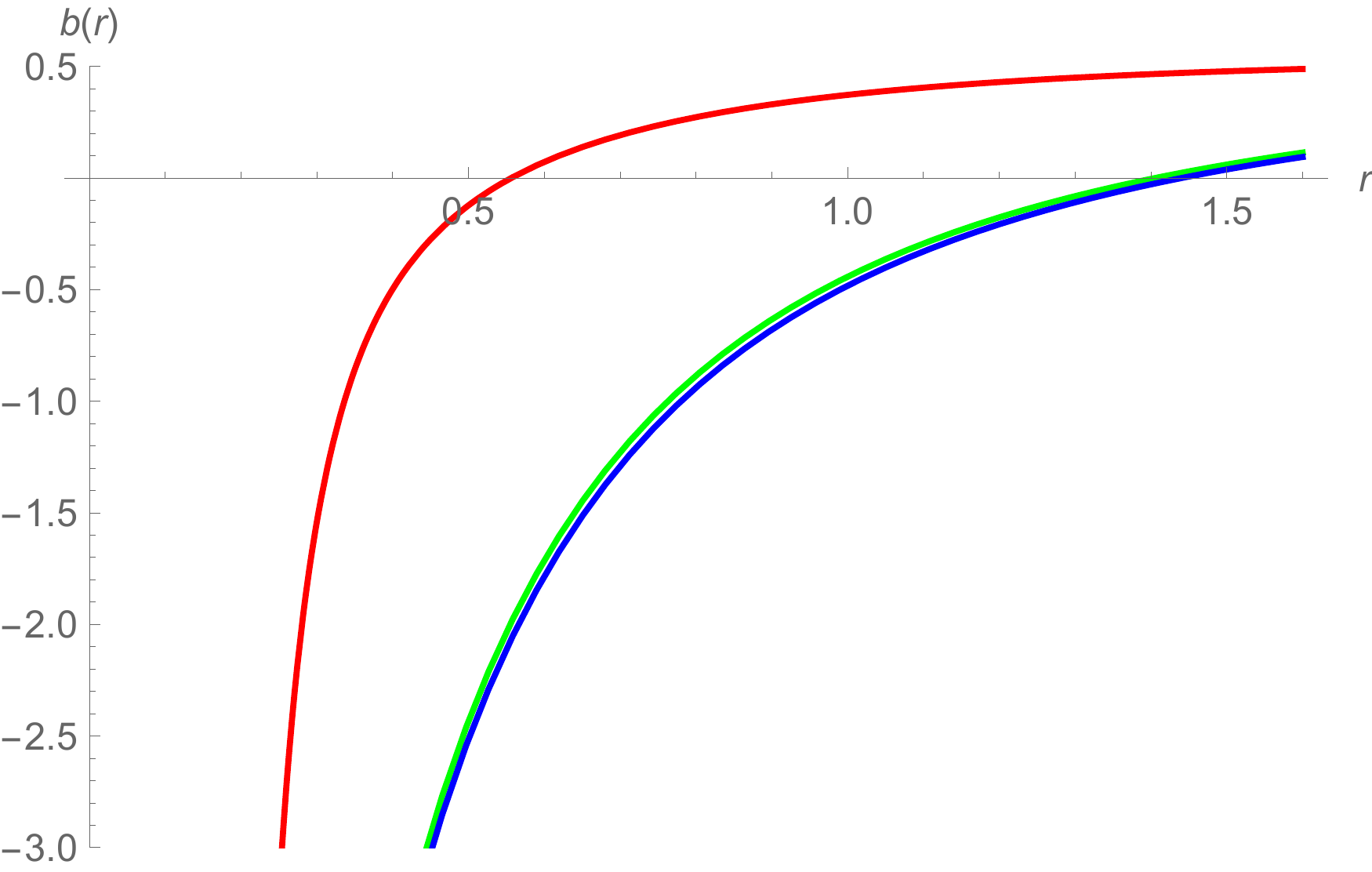}
\centering
\caption{$B(r)$ for $m=C_1=C_2=1$, $n=1$ (Green), $n=0.01$ (Red), $n=-0.05$ (Blue).}\label{fig1}
\end{figure}

\begin{figure}[h]
\includegraphics[width = 8cm, height = 4cm]{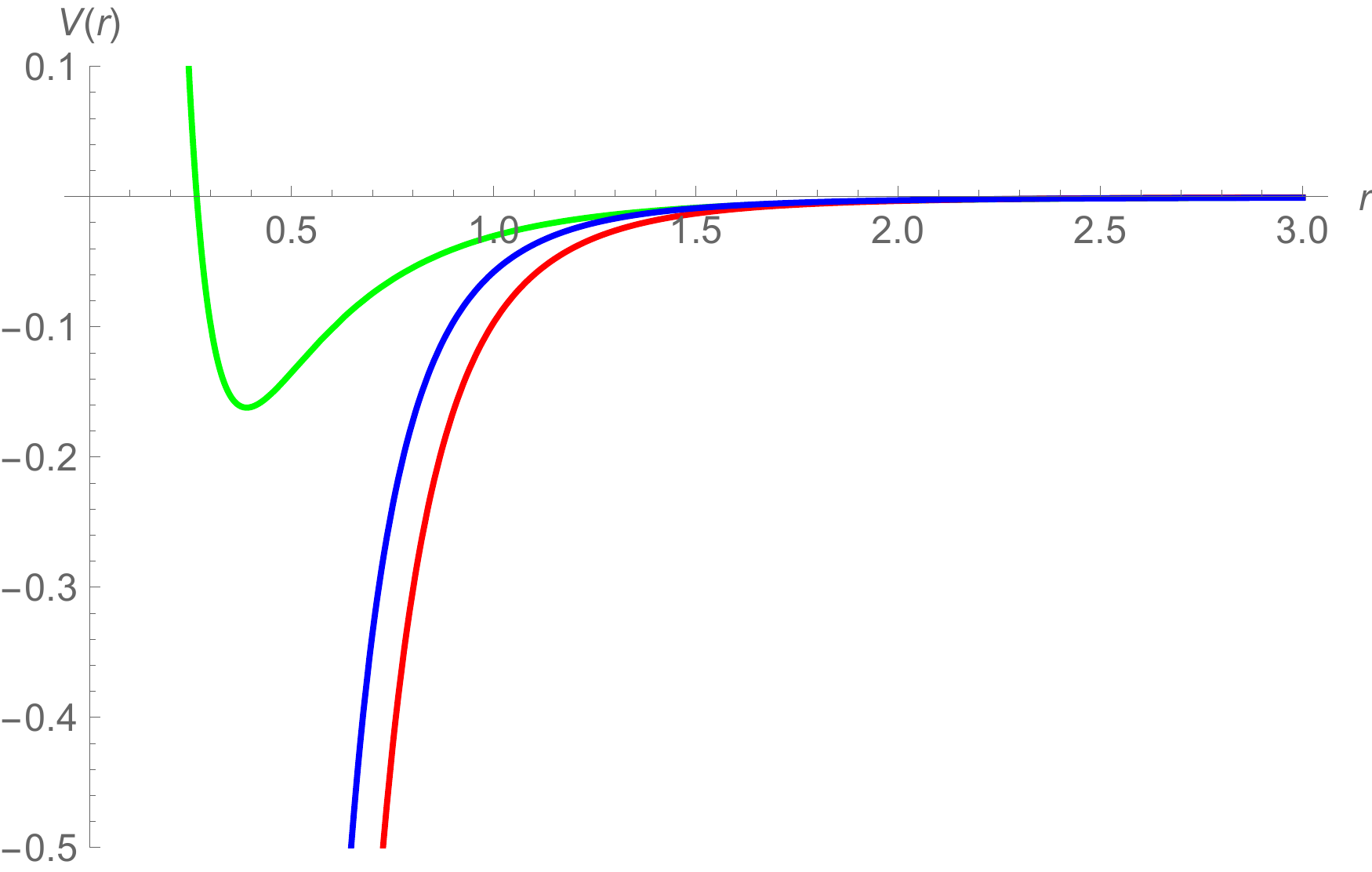}
\centering
\caption{$V(r)$ for $m=C_1=C_2=1$, $n=1$ (Green), $n=0.01$ (Red), $n=-0.05$ (Blue).}\label{fig2}
\end{figure}

The conditions we use here are: $B(0.2)=-10$, $B(100)=1$, $V(100)=0$.
We can see that in the all cases with the conditions we give, there is an horizon formed and  black hole solutions exist with a scalar field to behave  well for all $r>0$ values.

For the $f(R)$ function we have from equation (\ref{R1}),
\begin{eqnarray}
f'(R)&=&C_1+C_2 r(R)~,\\
f(R)&=&C_1 R+C_2\int^R r(R)dR~,\label{central}
\end{eqnarray}
where the parameter $C_1$ is dimensionless  and $C_2$ have dimensions $\left[C_2\right]=L^{-1}$. Therefore if
$C_1=1$ and $C_2\neq0$  this solution can be considered as an extension of the Einstein gravity. The scalar charges play a role in determining the metric therefore  the matter distribution will influence the form in the final $f(R)$  through the Ricci scalar.

\end{subsection}

\begin{subsection}{ $\phi(r) = me^{-nr}$}

Next we consider the case of the scalar field to be $\phi(r) = me^{-nr}~,$
where $m$ and $n$ are constants ($n>0$ for the appropriate asymptotic behaviour, with units $[L]^{-1}$). Then we get
\begin{equation} f_{R} =-\frac{1}{12} m^2 e^{-2 n r}+C_2 r+C_1~,\label{R2}
\end{equation}
and the metric function $B(r)$, $V(r)$ can be obtained numerically as shown in Fig. \ref{fig3} and Fig. \ref{fig4}.

\begin{figure}[h]
\includegraphics[width = 8cm, height = 4cm]{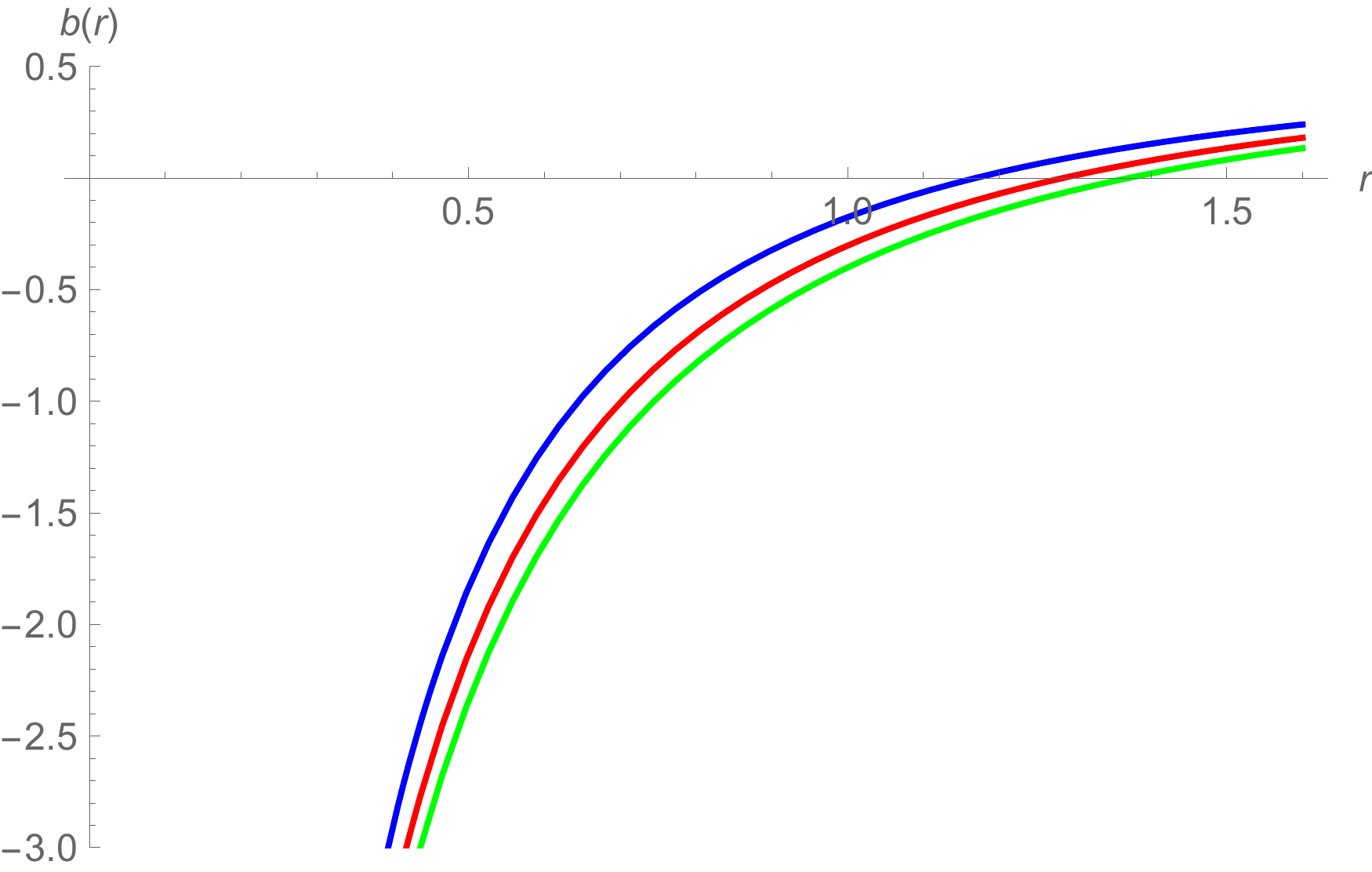}
\centering
\caption{$B(r)$ for $m=C_1=C_2=1$, $n=1$ (Green), $n=1.5$ (Red), $n=2$ (Blue).}\label{fig3}
\end{figure}
\end{subsection}

\begin{figure}[h]
\includegraphics[width = 8cm, height = 4cm]{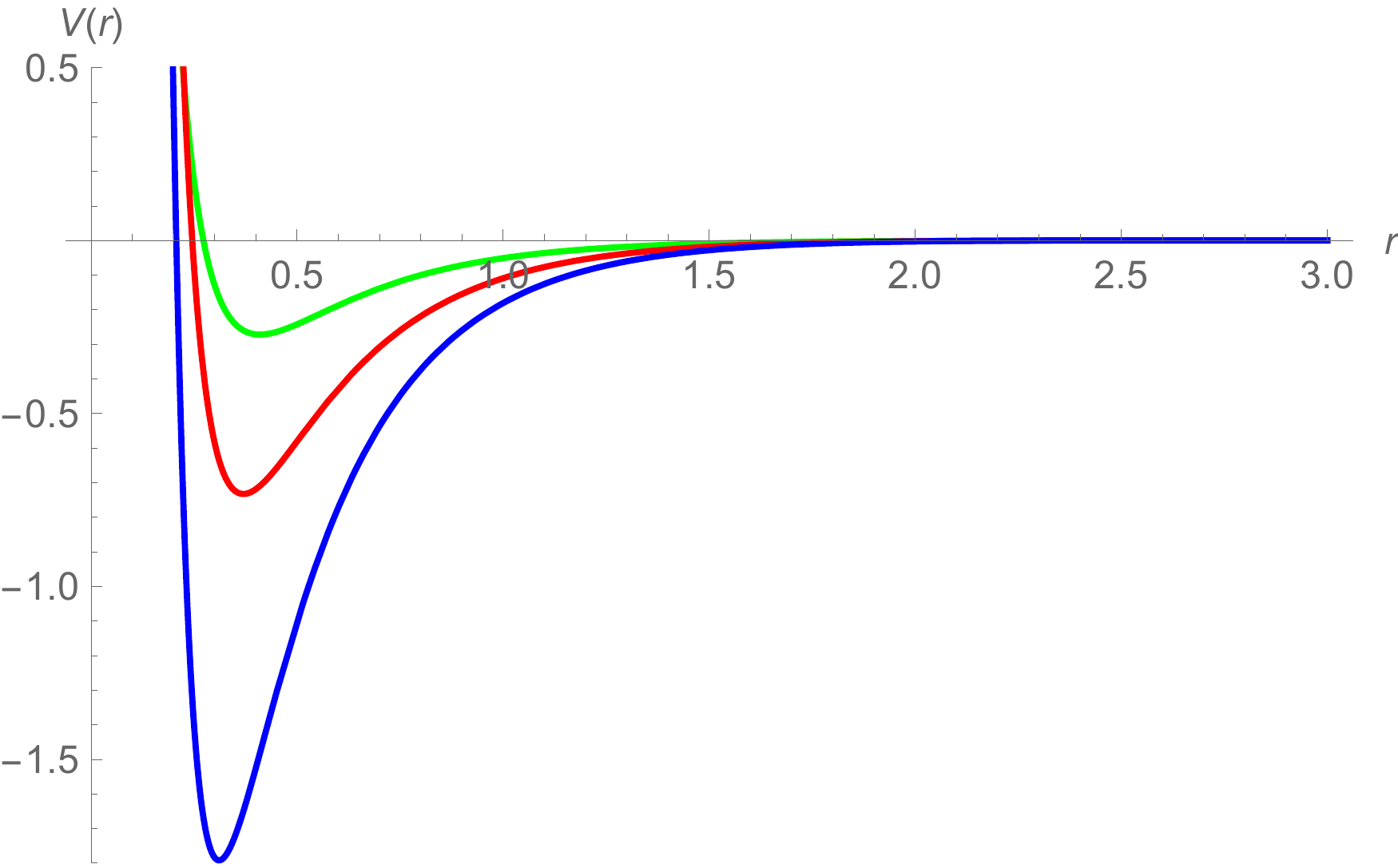}
\centering
\caption{$V(r)$ for $m=C_1=C_2=1$, $n=1$ (Green), $n=1.5$ (Red), $n=2$ (Blue).}\label{fig4}
\end{figure}

We can see that in all cases there is an horizon formed and  black hole solutions exist with a scalar field to behave well for all $r>0$ values.

For the $f(R)$ function we can see from equation (\ref{R2}) that curvature corrections will be present in the final $f(R)$ form. The first term of equation  (\ref{R2}) is related directly to the scalar field charges, while the last two terms will contain information about the scalar field indirectly through the metric function.

\newpage

\begin{subsection}{$\phi(r) = ArcTanh(\cfrac{m}{r+m}\big)$}

Finally we consider the case of the scalar field to be $$\phi(r) = ArcTanh(\cfrac{m}{r+m}\big)~,$$
where $m$ is a constant with units $[L]$ and we get
\begin{equation}f_{R} = \frac{(m+r) }{4 m}\ln\big(\frac{r}{2 m+r}\big)+\frac{1}{6} \tanh ^{-1}\left(\frac{m}{m+r}\right)^2+C_2 r+C_1~,
\label{R3} \end{equation}

and the functions $B(r)$, $V(r)$, $f_R$ and $\phi(r)$ can be obtained numerically as shown in Fig. \ref{fig5}-Fig. \ref{fig8}.

\begin{figure}[h]
\includegraphics[width = 8cm, height = 4cm]{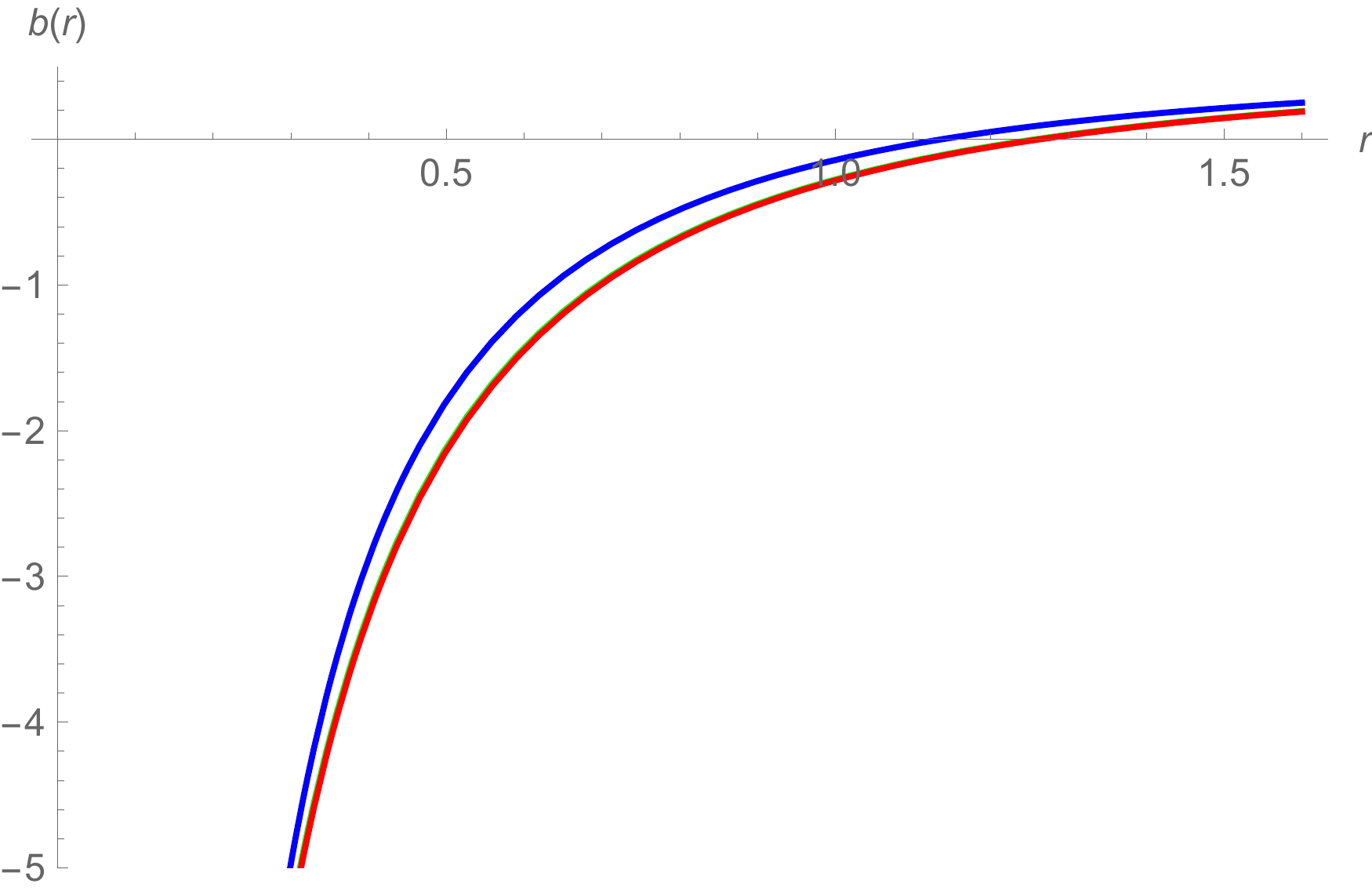}
\centering
\caption{$B(r)$ for $C_1=C_2=1$,$m=1$ (Green), $m=0.05$ (Red), $m=5$ (Blue) (Green and Red are very close).}\label{fig5}
\end{figure}

\begin{figure}[h]
\includegraphics[width = 8cm, height = 4cm]{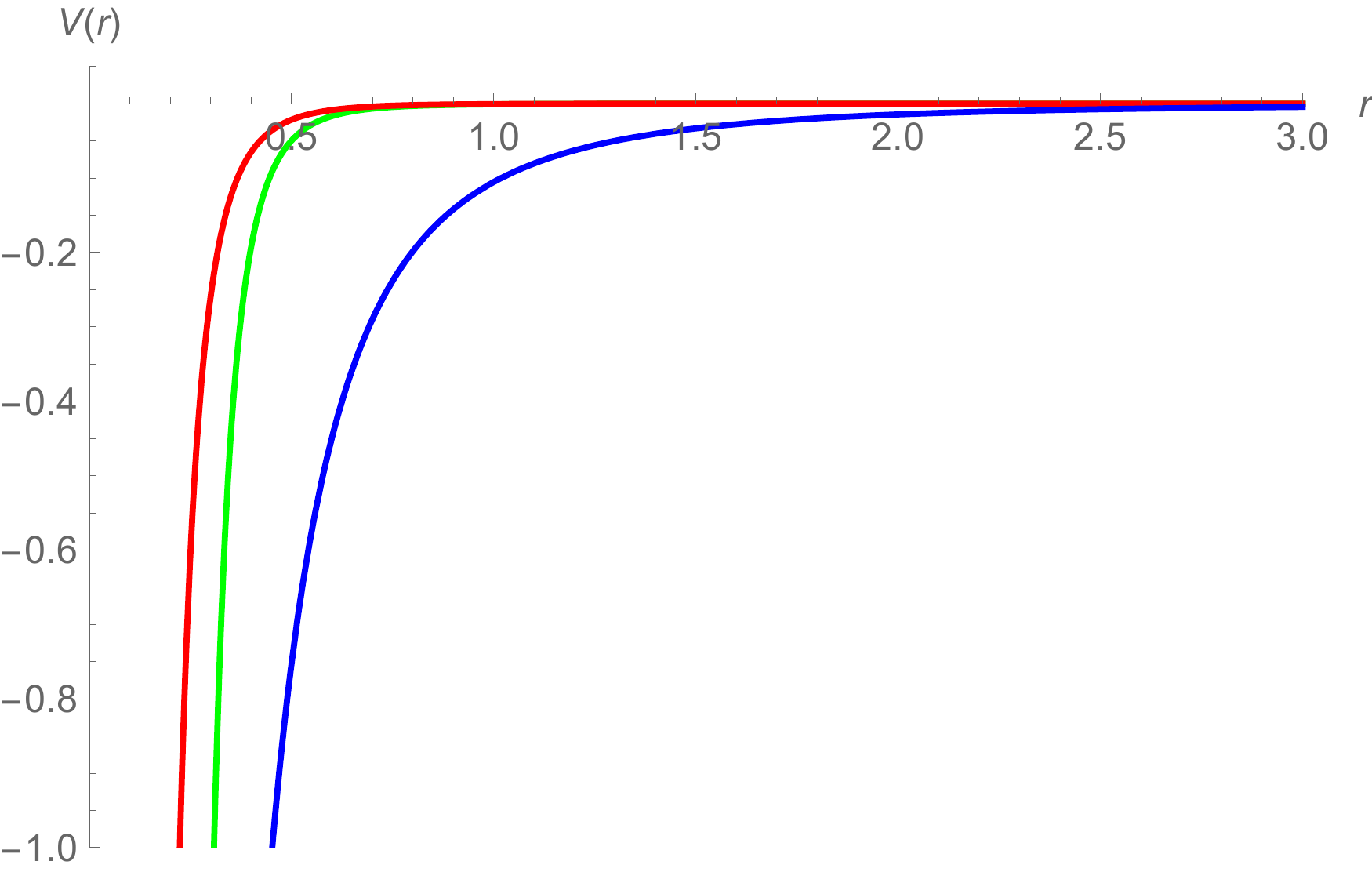}
\centering
\caption{$V(r)$ for $C_1=C_2=1$, $m=1$ (Green), $m=0.05$ (Red), $m=5$ (Blue).}\label{fig6}
\end{figure}

\begin{figure}[h]
\includegraphics[width = 8cm, height = 4cm]{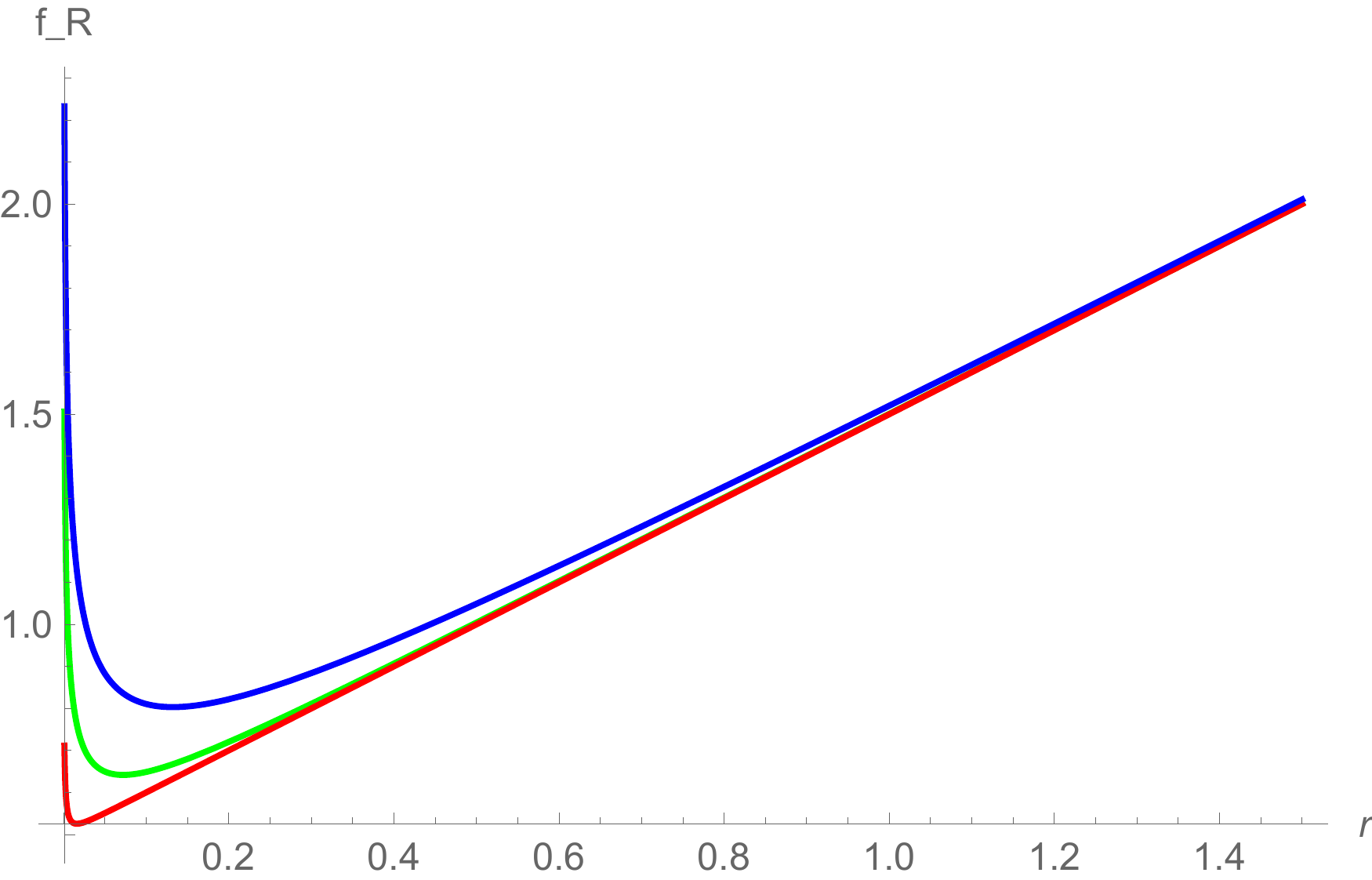}
\centering
\caption{$f_R$ for $C_1=C_2=1$, $m=1$ (Green), $m=0.05$ (Red), $m=5$ (Blue).} \label{fig7}
\end{figure}

\begin{figure}[h]
\includegraphics[width = 8cm, height = 4cm]{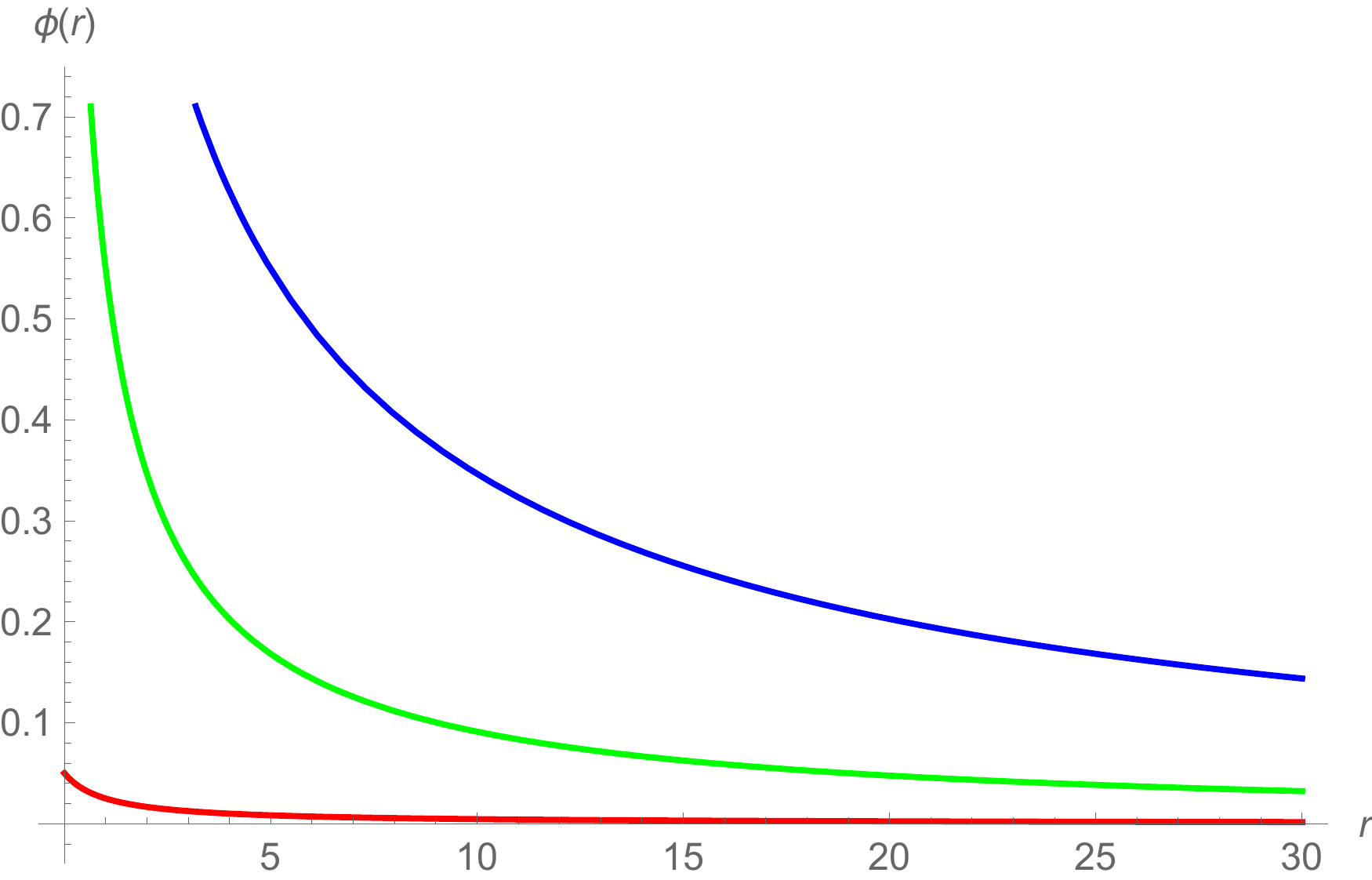}
\centering
\caption{$\phi(r)$ for $C_1=C_2=1$, $m=1$ (Green), $m=0.05$ (Red), $m=5$ (Blue).}\label{fig8}
\end{figure}
\end{subsection}

We can see that in all cases there is an horizon formed and  black hole solutions exist with a scalar field to behave  well for all $r>0$ values.

For the $f(R)$ function we can see from equation (\ref{R3}) that non-linear curvature corrections will appear in the final $f(R)$ form. These corrections are related directly to the scalar field charges due to equation (\ref{ttrr}) (the first two terms of equation (\ref{R3})). Of course, the scalar field plays a role in the metric function $B(r)$ so it is expected that information about the scalar field will  be present in the Ricci scalar and more non-linear corrections will finally appear because of the last two terms of equation (\ref{R3}), assuming of course that the Ricci scalar is dynamical.

We can see that in all the above cases, the scalar field modifies the gravitational model at hand, depending each time on the scalar field profile. The first polynomial distribution for example, seems to modify indirectly the gravitational model, while the other two distributions play a profound role in the final $f(R)$ model. The integration constants $C_1$ and $C_2$ have a physical meaning. $C_1$ is related to Einstein Gravity and $C_2$ is related to geometric corrections that can be encoded in $f(R)$ gravity, as it can be seen in equation (\ref{central}). Also observe that in all cases the potential well is formed before the formation of the horizon of the black hole which agrees with our findings in the case of a scalar field minimally coupled to gravity.

\section{Discussion and Conclusions}
\label{conclusions}

In this work we studied the production of scalarized black hole solutions in $f(R)$ gravity. Without specifying the form of the $f(R)$ function we introduced a scalar field minimally coupled to gravity. Solving the coupled field equations  we showed that there are no any black hole solutions. Introducing a self-interacting scalar potential and considering a matter distribution backreacting on the metric we found that the Schwarzschild-AdS black hole is scalarized at large distances depending on particular choices of parameters. With a different choice of parameters scalarized  Schwarzschild-AdS-like black holes can be found. The AdS space is generated by an effective cosmological constant which depends on the parameters.

At small distances we solved numerically the Einstein and Klein-Gordon equations. We found that the curvature is divergent at origin  $r \rightarrow 0$ indicating a singularity and all the functions are continuous for any positive $r$. The metric function develops an event horizon while the scalar potential  appears with  a deep well which is formed before the formation of the event horizon and also  develops a peak outside the event horizon. At small distances we find a similar behaviour of  a   scalar field conformally coupled to gravity. Considering  various matter distributions we found that at small distances hairy black holes are produced and the scalar potential develops a deep well before  the formation of the event horizon.

We attribute the scalarization of the black hole at large distances and the production of hairy black holes at small distances to pure curvature effects.
This is a result of the solutions of the field equations which introduce  non-linear curvature correction terms which appeared in the final $f(R)$ form. These correction terms  are related directly to the scalar field charges of the considered matter distributions. At large distances these correction terms are weak and the usual Ricci scalar term dominates and the coupling of the scalar field to the  metric scalarized the Schwarzschild-AdS or Schwarzschild-AdS-like black holes. At small distances the non-linear curvature correction terms dominate and then strong curvature and scalar dynamics give a rich structure of hairy black hole solutions.  We expect that matter is trapped in the formed  potential well and the scalar dynamics scalarized the black hole while after the formation of the hairy black hole it manifests itself with a peak of its potential outside the event horizon of the black hole. With a specific choice of parameters we found that the metric function exhibits a continuation from small to large distances connecting the solutions found in these regimes.

It would be very interesting to study in details the dynamical mechanism of how the matter trapped in the potential results to the formation of the hairy black hole. We have to stress here that  the mechanism of the formation of hairy black hole is a result of the direct coupling of matter to gravity.   There are other mechanisms of the formation of hairy black holes. On of them is the well known Gubser mechanism \cite{Gubser:2005ih} on which the  gauge/gravity holographic duality depends on it. A charged scalar field in the vicinity of a charged black hole in the AdS space is trapped outside the horizon of the black hole as a result of the competing forces of electromagnetic and gravitational forces. Another mechanism was discussed in \cite{Sotiriou:2014pfa,Doneva:2017bvd,Antoniou:2017acq}  and in its charged version \cite{Doneva:2018rou} in which a black hole is scalarized because matter is coupled to the Gauss-Bonnet term, which is a high order curvature term. We note here that our mechanism is different from this mechanism because the scalar field is not directly coupled to curvature but it feels the strong curvature effect only through its coupling to the metric.

\section{Acknowledgements}

We thank Ricardo Troncoso for his stimulating comments and suggestions.


\end{document}